\documentclass[useAMS,usenatbib]{mn2e}
\usepackage[pdftex]{graphicx}
\usepackage[pdftex,usenames,dvipsnames]{xcolor}
\usepackage{natbib}
\usepackage{aas_macros}
\usepackage{url}
\usepackage[none]{hyphenat}
\usepackage{times}
\usepackage[linktocpage=true,backref=true,pagebackref=false,bookmarks=true,bookmarksnumbered=False,colorlinks=true,linkcolor=NavyBlue,citecolor=NavyBlue,urlcolor=NavyBlue]{hyperref}
\usepackage[all]{hypcap}
\pdfpageattr {/Group << /S /Transparency /I true /CS /DeviceRGB>>}  

\title[Radio jet of IMBH]{The powerful jet of an off-nuclear intermediate-mass black hole in the spiral galaxy NGC 2276}

\author[M. Mezcua et al.]
  {M.~Mezcua,$^{1,2,3}$\thanks{Email: mar.mezcua@cfa.harvard.edu}
  T.P.~Roberts,$^{4}$
  A.P.~Lobanov$^5$\thanks{Visiting Scientist, University of Hamburg / Deutsches Elektronen Synchrotron Forschungszentrum.}
  A.D.~Sutton,$^{4}$\\
  $^1$Harvard-Smithsonian Center for Astrophysics (CfA), 60 Garden Street, Cambridge, MA 02138, USA \\
  $^2$Instituto de Astrof\'isica de Canarias (IAC), E-38200 La Laguna, Tenerife, Spain \\
  $^3$Department Astrof\'isica, Universidad de La Laguna, E-38206 La Laguna, Tenerife, Spain \\ 
  $^4$Department of Physics, University of Durham, South Road, Durham DH1 3LE, UK \\
  $^5$Max Planck Institute for Radio Astronomy, Auf dem H\"ugel 69, D-53121 Bonn, Germany }

\date{Accepted 2015 January 20}

\pagerange{\pageref{firstpage}--\pageref{lastpage}} \pubyear{2013}

\def\LaTeX{L\kern-.36em\raise.3ex\hbox{a}\kern-.15em
    T\kern-.1667em\lower.7ex\hbox{E}\kern-.125emX}

\begin{document}

\label{firstpage}

\maketitle

\begin{abstract}

Jet ejection by accreting black holes is a mass invariant mechanism unifying stellar and supermassive black holes (SMBHs) that should also apply for intermediate-mass black holes (IMBHs), which are thought to be the seeds from which SMBHs form. We present the detection of an off-nuclear IMBH of $\sim$5 $\times$ 10$^{4}$ M$\odot$ located in an unusual spiral arm of the galaxy NGC 2276 based on quasi-simultaneous \textit{Chandra} X-ray observations and European VLBI Network (EVN) radio observations. The IMBH, NGC2276-3c, possesses a 1.8 pc radio jet that is oriented in the same direction as large-scale ($\sim$650 pc) radio lobes and whose emission is consistent with flat to optically thin synchrotron emission between 1.6 and 5 GHz. Its jet kinetic power ($4 \times 10^{40}$ erg s$^{-1}$) is comparable to its radiative output and its jet efficiency ($\geq$ 46\%) is as large as that of SMBHs. A region of $\sim$300 pc along the jet devoid of young stars could provide observational evidence of jet feedback from an IMBH. The discovery confirms that the accretion physics is mass invariant and that seed IMBHs in the early Universe possibly had powerful jets that were an important source of feedback.
\end{abstract}


\section{Introduction}
Intermediate-mass black holes (IMBHs) constitute the missing link between stellar-mass and supermassive black holes (SMBHs) and are the potential long-sought seeds from which SMBHs grow. IMBHs could form either from the death of very massive and short-lived stars, the direct collapse of a pre-galactic gas disc, or the collapse of dense stellar clusters, and then grow either through hierarchical merging or secular gas accretion (\citealt{2010A&ARv..18..279V}). They should thus be present in the nuclei of low-mass galaxies, as predicted from BH mass ($M_\mathrm{BH}$)-scaling relations (e.g. \citealt{2013ApJ...764..151G}); in dense star-forming regions in spiral galaxies; and in the haloes of large galaxies (e.g.  after tidal stripping of merging low-mass satellite galaxies).
SMBHs with masses up to 10$^{9}$ M$\odot$ already existed when the Universe was less than $\sim$1 Gyr old (e.g. \citealt{2011Natur.474..616M}). To reach this mass in such a short time, seed IMBHs must have undergone short phases of growth above the Eddington rate coupled with enhanced star formation triggered by jets or outflows (\citealt{2014arXiv1401.3513V}) or supra-exponential accretion (\citealt{2014Sci...345.1330A}).
Cosmological simulations show that the radiative and mechanical feedback from an accreting IMBH regulate its growth and deprive the inner core of the galaxy of cold star-forming gas (\citealt{2012MNRAS.420.2662D}; \citealt{2011ApJ...738...54K}). However, the detection of IMBHs and, in particular, of jet emission from IMBHs is scarce and no evidence of its feedback effects has hitherto been observed. 

The best observational evidence for IMBHs has been found in the nuclei of low-mass spiral and dwarf galaxies (\citealt{2007ApJ...670...92G}; \citealt{2007ApJ...657..700D}; \citealt{2008ApJ...686..892T}; \citealt{2013ApJ...775..116R}; \citealt{2013ApJ...773..150S}; \citealt{2014ApJ...782...55Y}) and in ultraluminous X-ray sources (ULXs; \citealt{2009Natur.460...73F}). Evidence is mounting that the high X-ray luminosity of the majority of ULXs can be explained by stellar-mass BHs accreting at around or above the Eddington limit (e.g.  \citealt{2009MNRAS.397.1836G}; \citealt{2013MNRAS.435.1758S}; \citealt{2014Natur.514..198M}). However, those ULXs with L$_\mathrm{X} > 5\times10^{40}$ erg s$^{-1}$ remain difficult to explain as super-Eddington accretion on to stellar-mass BHs, and display X-ray emission properties consistent with IMBHs either accreting in a sub-Eddington hard state (\citealt{2012MNRAS.423.1154S}) or in an intermediate state (\citealt{2014Natur.513...74P}). One such object is a ULX in NGC 2276 (NGC2276-3c for simplicity), which had a peak X-ray luminosity of up to $\sim 6\times10^{40}$ erg s$^{-1}$ when observed with the \textit{XMM-Newton} satellite (see Section~\ref{chandra} for further details). If they are in the sub-Eddington hard or intermediate states, accreting BHs are expected to emit radio jets. Observational evidence for compact or transient jet radio emission has been found in a few ULXs (\citealt{2011AN....332..379M}; \citealt{2012ApJ...749...17C,2014MNRAS.439L...1C}; \citealt{2012Sci...337..554W}; \citealt{2013Natur.493..187M}; \citealt{2013MNRAS.436.1546M,2014ApJ...785..121M}). The best evidence for the presence of a radio jet in an off-nuclear IMBH comes from ESO 243-49 HLX-1, which shows flaring radio emission when appearing to transition from the low/hard X-ray state to the high/soft X-ray state (\citealt{2012Sci...337..554W}; \citealt{2015MNRAS.446.3268C}). However, observational evidence of extended, steady jet radio emission from IMBHs has so far only been found in the nuclear IMBHs NGC 4395 (jet extent $\sim$0.3pc; \citealt{2006ApJ...646L..95W}) and GH10 (jet extent $<$ 320 pc; \citealt{2008ApJ...686..838W}).
For the ULX NGC2276-3c, extended radio emission formed by two lobes of total size $\sim$650 pc was detected with the Karl G. Jansky Very Large Array (VLA). The X-ray source is located in between the two radio lobes, suggesting the presence of a central BH powering a two-sided radio jet (\citealt{2013MNRAS.436.3128M}). Using the X-ray peak luminosity, a $M_\mathrm{BH} \geq$ 4700 M$\odot$ was estimated assuming 10\% Eddington accretion (\citealt{2012MNRAS.423.1154S}). The VLA observations could thus have revealed the largest radio jet ever detected from an IMBH. 

In this paper, we report quasi-simultaneous very long baseline interferometry (VLBI) radio observations with the European VLBI Network (EVN) and X-ray observations with the \textit{Chandra} X-ray observatory of NGC2276-3c, which reveal the detection of a parsec-scale radio jet with an orientation consistent with that of the VLA radio lobes and allow us to estimate a $M_\mathrm{BH}$ in the IMBH range. 
A description of the observations and results is provided in Section~\ref{observations}. The discussion of the results and final conclusions are given in Sections~\ref{discussion} and \ref{conclusions}, respectively.

\begin{figure*}
 \includegraphics[width=\textwidth]{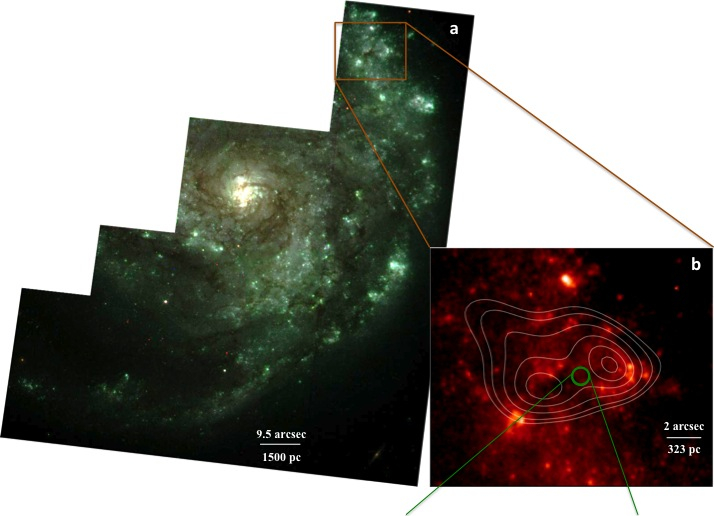}
  \includegraphics[width=1.015\textwidth]{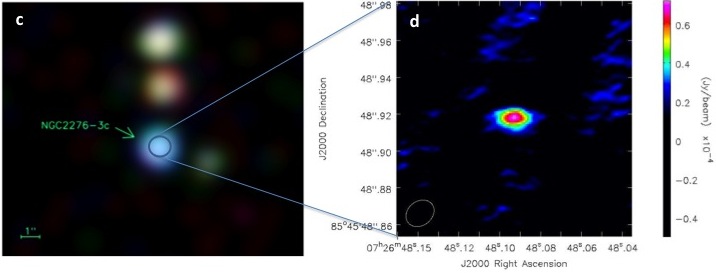}
\caption{\textbf{a)} Three-colour image of the western arm of NGC\,2276 where the ULX is located. The image has been created using three filters of the Wide Field Planetary Camera 2 on the \textit{HST}: \textit{F814W} in red ($\sim$0.8 $\mu$m), \textit{F606W} in green ($\sim$0.6 $\mu$m) and \textit{F550W} in blue ($\sim$0.55 $\mu$m). \textbf{b)} \textit{I}-band ($\sim$0.8 $\mu$m) \textit{HST} image of the region where the ULX NGC2276-3c is located. The 5 GHz VLA radio lobes \citep{2013MNRAS.436.3128M} are overplotted as (5,6,7,8,9,10) times the off-source rms noise of 0.04 mJy beam$^{-1}$. The \textit{Chandra} X-ray position is marked with a green circle of $\sim$1 arcsec diameter. The VLA beam size is 4.6 arcsec $\times$ 3.5 arcsec. \textbf{c)} \textit{Chandra} red (0.2--1.5 keV), green (1.5--2.5 keV) and blue (2.5--8 keV) image convolved with an $\sim$2 arcsec FWHM Gaussian. The position of NGC2276-3c is marked with a blue circle of $\sim$1 arcsec diameter. \textbf{d)} EVN 1.6 GHz image of NGC2276-3c. The synthesized beam size is 16.4 $\times$ 13.1 mas oriented at a P.A. of $-52^{\circ}$. The off-source rms noise is 8 $\mu$Jy beam$^{-1}$.  }
\label{fig1}
\end{figure*}

\section{Observations, analysis and results}
\label{observations}
\subsection{EVN observations}
EVN observations of NGC2276-3c were carried out during two 6-h observing runs on 2013 May 23 and June 3 at 5 and 1.6 GHz, respectively (experiments EM106A and EM106B). The observations were carried out in the phase-reference mode, alternating scans between the target source and a nearby ($\Delta\theta\sim3^{\circ}$.2) compact source used as phase calibrator (J0508+8432). In total, $\sim$3 h were spent on NGC2276-3c at both frequencies. The bright radio source 4C+39.25 was also observed to be used as fringe finder and bandpass calibrator. The data were recorded at a rate of 1024 Mbps in dual circular polarization, using eight sub-bands (each 16 MHz in width and split into 32 spectral channels), and correlated at the EVN Correlator Facility of the Joint Institute for VLBI in Europe (JIVE) with an averaging time of 1 s. Amplitude calibration (using system temperatures and antennas gains) and fringe fitting was performed using the \textsc{AIPS} software. Ionospheric calibration was also applied at both frequencies.
Radio maps at 1.6 GHz (Fig.~\ref{fig1} d) and 5 GHz were obtained using hybrid imaging with \textsc{CLEAN} deconvolution and the natural weighting of the data. At 1.6 GHz, radio emission at a 9$\sigma$ level is detected in between the two peaks of the radio lobes previously detected with the VLA (Fig.~\ref{fig1}, b). The emission has a flux density of 65 $\pm$ 16 $\mu$Jy, from which we derive an integrated radio luminosity $L_\mathrm{1.6 GHz}$ = 1.4 $\times$ 10$^{35}$ erg s$^{-1}$ assuming a distance to NGC 2276 of 33.3 Mpc (\citealt{2012MNRAS.423.1154S}). This spectral luminosity is of the same order as that of the IMBH in NGC 4395 (\citealt{2006ApJ...646L..95W}) and one order of magnitude lower than the radio luminosity of parsec-scale jets in active galactic nuclei (AGN) residing in spiral galaxies (e.g. NGC 1097; \citealt{2014ApJ...787...62M}). 
The emission was fitted by a two-dimensional elliptical Gaussian component, which is centred at RA(J2000) = 07$^{h}$26$^{m}$48$^{s}$.092 $\pm$ 0$^{s}$.001, Dec.(J2000) = +85$^{\circ}$45\arcmin 48\arcsec.9180 $\pm$ 0\arcsec.0006, has a major axis deconvolved size of 11.3 mas ($\sim$1.8 pc) oriented at a position angle of 78$^{\circ}$ and a brightness temperature $T_\mathrm{B}$ = 1.5 $\times$ 10$^{5}$ K. The deconvolved size of 11.3 mas is larger than the resolution limit\footnote{The resolution limit is given by 
$\theta_\mathrm{lim} = \frac{2}{\rm{\pi}}\left[\rm{\pi} b_\mathrm{maj} b_\mathrm{min}\ln\left((\rm{SNR}+1)/\rm{SNR}\right)\right]^{1/2}$, where $b_\mathrm{maj}$ and $b_\mathrm{min}$ are the beam major and minor axis, respectively, and SNR the signal-to-noise ratio (\citealt{2005astro.ph..3225L}).} of 5.4 mas, implying that the emission is resolved along the major axis. Its orientation, at a position angle of 78$^{\circ}$ north-through-east, is consistent with that of the VLA radio lobes. The finding of $T_\mathrm{B} > 10^{5}$ rules out thermal free--free emission (of typically 10$^{4}$ K) from e.g.  H\textsc{II} regions, suggesting that the emission may be non-thermal. The same result is obtained when performing hybrid imaging with \textsc{CLEAN} deconvolution plus self-calibration loops, which proves that the extended emission is not caused by phase errors.

The radio map at 5 GHz was produced using the data from baselines shorter than 20 M$\lambda$ (1200 km) in order to improve the detection of extended emission. A restoring beam of 16.4 $\times$ 13.1 mas full width at half-maximum (FWHM) was used in order to match that of the 1.6 GHz radio map and derive the spectral index. A compact component of flux density 39 $\pm$ 18 $\mu$Jy and radio luminosity of $L_\mathrm{5 GHz}$ = 2.6 $\times$ 10$^{35}$ erg s$^{-1}$ is detected at the $\sim$~8$\sigma$ level. The component is centred at RA(J2000) = 07$^{h}$26$^{m}$48$^{s}$.093 $\pm$ 0$^{s}$.001, Dec.(J2000) = +85$^{\circ}$45\arcmin 48\arcsec.912 $\pm$ 0\arcsec.001. It has a size of 13.3 $\times$ 12.7 mas oriented at a position angle of 11$^{\circ}$ and $T_\mathrm{B}$ = 1.1 $\times$ 10$^{4}$ K. The total positional errors of the target source are estimated at both frequencies as a quadratic sum of the positional error of the target source in the phase-referenced map, the positional error of the phase-reference calibrator and the error of phase referencing due to ionospheric effects. This yields a total positional error for NGC2276-3c of 3.5 mas at 1.6 GHz and 2.4 mas at 5 GHz. The position of the components identified at each frequency coincides within these positional errors. A spectral index $\alpha$ = $-0.5\ \pm$ 0.2, S$_{\nu} \propto \nu^{\alpha}$, is derived from the integrated flux densities at 1.6 and 5 GHz. Using the peak densities, an even flatter spectral index $\alpha$ = $-0.3\ \pm$ 0.1 would be obtained. In both cases, the spectral index is consistent with flat to optically thin synchrotron emission.

\subsection{\textit{Chandra} observations}
\label{chandra}
A \textit{Chandra} X-ray observation (ID 15648) was carried out
quasi-simultaneously with (i.e. started within one day of the
completion of) the 5-GHz radio observations, with the target placed at the
nominal aim-point on the S3 chip of the ACIS-S array. The source was
observed for an exposure time of 25 ks, and the observation was
processed using standard tools in CIAO v4.6, with calibration files
from CALDB v4.6.1.1. In total, 188 counts were detected from the ULX,
yielding a 0.3--10 keV count rate of (7.5 $\pm$ 0.6) $\times$ 10$^{-3}$ count s$^{-1}$. The source was also previously detected by \textit{Chandra}
on 2004 June 23 (ID 4968), when it had a count rate of $\sim2.6 \times$
10$^{-3}$ count s$^{-1}$ (\citealt{2011AN....332..358W}; \citealt{2012MNRAS.423.1154S}).
Hence, NGC2276-3c was $\sim 3$ times brighter in the new observation
than in the only previous observation in which it was resolved from
its neighbouring ULXs. 

We used XSPEC v12.8.1 to fit models to the binned X-ray spectrum. Both
an absorbed power-law and a multicolour-disc (MCD) model were used to
fit the data. In both cases, we included two absorption components:
the first was set equal to the line-of-sight Galactic column density
in the direction of NGC 2276 (5.52 $\times$ 10$^{20}$ cm$^{-2}$;
\citealt{1990ARA&A..28..215D}), whilst the other was free to vary, to model
any intrinsic absorption in the source and/or the host galaxy. The fit
statistics for the power-law and MCD were $\chi^2$/degrees of freedom
= 13.7/15 and 13.2/15 respectively. We extracted the best-fitting
model parameters from the absorbed power-law and MCD. These were
$N_\mathrm{H}$ = 1.3$^{+0.5}_{-0.4} \times 10^{22}$ cm$^{-2}$ and
$\Gamma = 1.4 \pm 0.3$ for the power-law, and $N_\mathrm{H}$ = (0.9 $\pm$
0.3) $\times$ 10$^{22}$ cm$^{-2}$ and $k_\mathrm{B}T_\mathrm{in}$ =
2.4$^{+1.0}_{-0.5}$ keV for the MCD. All of these parameters are
consistent with the values reported by \cite{2012MNRAS.423.1154S} for the
previous \textit{Chandra} observation, and only the normalizations differ
significantly. The parameters are also consistent, within the errors, with those obtained when fitting the unbinned data using the C-statistic in {\sc xspec} for Poisson data with a Poisson background\footnote{https://heasarc.gsfc.nasa.gov/xanadu/xspec/manual/XSappendixStatistics.html}. Interestingly the high disc temperature from the MCD
fit is similar to that seen in Galactic black holes in the steep
power-law state (bright intermediate states), and so appears somewhat
too high for an IMBH.  However, we note that the moderate data quality may
simply be insufficient to reject an MCD fit to a power-law spectrum in
this case.

We estimated observed X-ray fluxes from the absorbed power-law model,
which we converted to luminosities. These are: $L_{\rm 0.3-10~keV} =
1.8^{+0.3}_{-0.2} \times 10^{40}~{\rm erg~s^{-1}}$ and $L_{\rm
2-10~keV} = 1.6^{+0.3}_{-0.2} \times 10^{40}~{\rm erg~s^{-1}}$. This
corresponds to an increase in luminosity of $\sim 3.5$ from the 2004
{\it Chandra} observation. Although NGC2276-3c has increased in X-ray
luminosity, the total luminosity of this and the other nearby point
sources (see below) is still less than that of the extreme luminosity
ULX observed by {\it XMM--Newton} (2XMM J072647.9$+$854550; 0.3--10 keV
luminosity of $(6.1 \pm 0.3) \times 10^{40}~{\rm erg s^{-1}}$), meaning
that at least one of the point sources detected by {\it Chandra}
within the {\it XMM-Newton} beam must have previously been even more
luminous in X-rays. Given the location of the centroid of the {\it
XMM--Newton\/} detection (cf. Fig. 1 of \citealt{2012MNRAS.423.1154S}), it is
likely that the brighter source was either 3b or 3c.

We also examined the {\it Chandra\/} light curve of the ULX to test
for evidence of variability. However, the 3$\sigma$ upper limit on the
rms fractional variability is very unconstrained at $\sim
0.6$. Although the data are relatively poor, we note that the
characteristics of a hard accretion state -- a power-law spectral form
with $\Gamma \sim 1.7$ and rms variability at the 10-20 per cent
level -- are well within the range of allowed parameters we have
derived. Given the detection of a radio jet characteristic of the hard
state, a thermal-dominant state appears unlikely, although we cannot
statistically rule out an accretion disc-like spectrum (but we again
note that the derived temperature is somewhat higher than typical for
a thermal dominant state). We can therefore conclude that the data are
consistent with a hard accretion state.

Motivated by the large change in the X-ray luminosity of NGC2276-3c,
we also tested whether the neighbouring sources (3a and 3b in \citealt{2012MNRAS.423.1154S}) had varied too. 
To do this, we extracted energy spectra from the latest {\it Chandra\/} observation using the same method as
above. We then fitted both of the ungrouped spectra with a doubly
absorbed power law, by minimizing the modified Cash-statistic (which
allows for a background spectrum) in {\sc xspec}. We find that source
3a is well modelled by a power-law spectrum with $N_{\rm H} = (5 \pm
2) \times 10^{21} \rm ~cm^{-2}$, $\Gamma \sim 1.7 \pm 0.2$ and
observed 0.3--10\,keV luminosity $L_{\rm X} = (1.3 \pm 0.2) \times
10^{40} \rm ~erg~s^{-1}$, whereas source 3b has $N_{\rm H} = (6 \pm 2)
\times 10^{21} \rm ~cm^{-2}$, $\Gamma = 2.2\pm 0.4$ and observed
0.3-10\,keV luminosity $L_{\rm X} = 5.5^{+1.1}_{-0.9} \times 10^{39}
\rm ~erg~s^{-1}$.  These model parameters and fluxes are consistent
(within the 90 per cent uncertainty regions) with the previous values
from the 2004 observation.

\section{Discussion}
\label{discussion}
\subsection{The nature of NGC2276-3c}
\begin{figure}
 \includegraphics[width=\columnwidth]{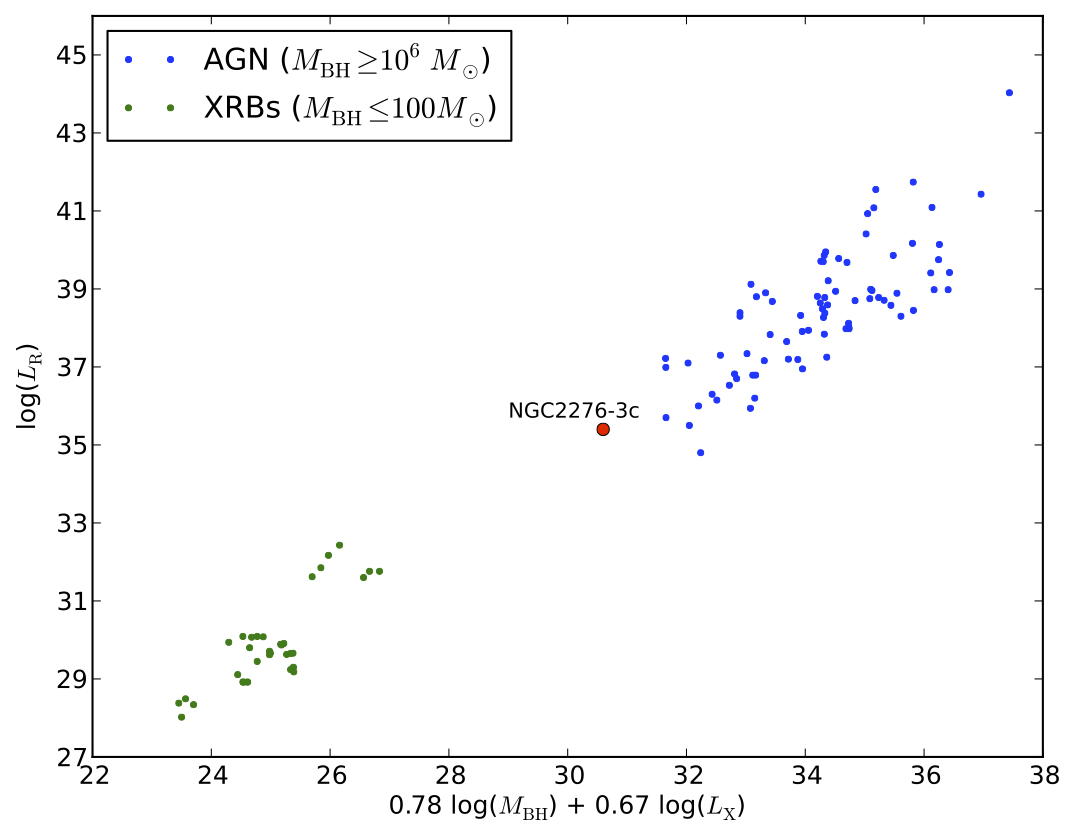}
\caption{Fundamental Plane of BH accretion from \citet{2009ApJ...706..404G} extending from XRBs hosting stellar-mass BHs (green dots) to AGN hosting SMBHs (blue dots). The sample of XRBs and AGN is taken from \citet{2003MNRAS.345.1057M}. A clear gap in the correlation is observed in the IMBH range, where NGC2276-3c is located.}
\label{fig2}
\end{figure}

The spatial coincidence of the EVN non-thermal radio emission with the location of unresolved, hard X-ray emission constitutes one of the most compelling radiative signatures of an accreting BH. It also permits a measurement of the $M_\mathrm{BH}$ using the Fundamental Plane of BH accretion, which is a correlation between core radio luminosity, X-ray luminosity and $M_\mathrm{BH}$ valid from stellar to SMBHs in the hard X-ray spectral state, where steady jet emission is ubiquitous. For this purpose, we use the empirical Fundamental Plane dependence of \citeauthor{2009ApJ...706..404G} (2009, eq. 3), which is the only one that has been tested in the IMBH range (\citealt{2014ApJ...788L..22G}). Using the EVN radio luminosity at 5 GHz and the \textit{Chandra} X-ray luminosity in the 2-10 keV band, we obtain $M_\mathrm{BH}$ = 5 $\times$ 10$^{4}$ M$\odot$ with a scatter of 0.7 dex, which confirms that the ULX harbours an IMBH (Fig.~\ref{fig2}). The mass obtained using other Fundamental Plane correlations (e.g. \citealt{2003MNRAS.345.1057M}; \citealt{2012ApJ...755L...1M}; \citealt{2012MNRAS.419..267P}) is also consistent with an IMBH, though we note that these have not been proven to be valid in the IMBH range and sometimes assume certain accretion physics that do not necessarily apply to NGC2276-3c. Further uncertainties also arise from the discovery of a second track in the Fundamental Plane (\citealt{2012MNRAS.423..590G}) and of no dependence of the radio/X-ray correlation with BH mass in X-ray binaries (XRBs; \citealt{2014MNRAS.445..290G}).

To further probe the nature of the source, we also calculate the $R_\mathrm{X}$ ratio of 5 GHz radio emission to 2--10 keV X-ray emission: $R_\mathrm{X}$ = $\nu L_{\nu}$(5 GHz)/$L_{X}$(2--10 keV) (\citealt{2003ApJ...583..145T}). XRBs have typical values of log $R_\mathrm{X} < -5.3$, low-luminosity AGN have $-3.8 <$ log $R_\mathrm{X} < -2.8$, and supernova remnants have log $R_\mathrm{X} \sim-2$ (e.g. \citealt{2013MNRAS.436.2454M}; see also table~3 in \citealt{2013MNRAS.436.1546M} and references therein). NGC2276-3c has log $R_\mathrm{X} = -4.9$, which falls in the range $-$5.3 to $-$3.8 expected for IMBHs (\citealt{2013MNRAS.436.1546M}) and rules out a SMBH nature. A background AGN is also disfavoured by the location of the ULX within a spiral arm in which the VLA radio lobes and the X-ray source appear to line up with a cavity in the optical emission (Fig.~\ref{fig1}, b), and by the low probability of a chance alignment with a background source calculated from the VLA field (\citealt{2013MNRAS.436.3128M}).  
It should be noted that the fact that this ULX has a large-scale (lobe-like) structure and a small-scale jet indicates that other ULX masses derived for similar sources using VLA flux density measurements may be biased towards higher values (since the radio luminosity obtained from VLA measurements can be overestimated).


\subsection{Eddington ratio and accretion rate}
The Fundamental Plane has been applied under the assumption that the source is accreting at a sub-Eddington rate. The Eddington ratio is given by $k_\mathrm{bol} L_\mathrm{X}/[1.38 \times 10^{38} \times (M_\mathrm{BH}/M_\odot)]$ where $k_\mathrm{bol}$ is the 2--10 keV bolometric correction factor and ranges from ~5 (\citealt{2012A&A...539A..48M}) to $\sim$30 for AGN with Eddington ratios $\leq$ 0.1 (\citealt{2009MNRAS.392.1124V}) and $k_\mathrm{bol}\sim1$ for XRBs (e.g.  \citealt{2005Ap&SS.300..197M}). Assuming $k_\mathrm{bol}=5$ for NGC2276-3c (i.e. in between those of AGN and XRBs), and using $M_\mathrm{BH}$ = 5 $\times$ 10$^{4}$ M$\odot$ and the 2-10 keV X-ray luminosity, yields an Eddington ratio $\sim10^{-2}$.
The possibility that NGC2276-3c is accreting at a higher Eddington rate should also be considered. If the source is accreting at the Eddington limit, it has $M_\mathrm{BH} \geq 725$ M$\odot$, which still puts NGC2276-3c in the IMBH range. Even in the most conservative scenario that all the luminosity is emitted in the 2-10 keV X-ray band  ($k_\mathrm{bol}=1$) and there is no intrinsic absorption, the source would have $M_\mathrm{BH} \geq 145$ M$\odot$. Therefore, NGC2276-3c would qualify as an IMBH even if it were accreting at its Eddington rate. However, we now know that ULXs can be powered by super-Eddington emission from a stellar-mass BH (e.g.  \citealt{2013Natur.493..187M}: \citealt{2013Natur.503..500L}). Given that the maximal emission from a super-Eddington accretion disc is a factor $\sim$20 above the Eddington luminosity (e.g.  \citealt{2011AN....332..402M}\footnote{This factor might be further exceeded in ULXs that are able to strongly beam their emission, for example magnetic neutron stars such as the object underlying M82-X2 (\citealt{2014Natur.514..202B}).}), in the case of NGC 2276-3c we would require a massive stellar BH ($>$ 31 M$\odot$) to power the peak X-ray luminosity. We note, though, that the presence of such a powerful, extended radio jet is unprecedented for a ULX accreting at super-Eddington rates. Empirical evidence that BHs with supercritical accretion can have powerful jets has so far only been found in two stellar-mass BHs (S26, in the nearby galaxy NGC\,7793, \citealt{2010Natur.466..209P}; and MQ1, in the spiral galaxy M83, \citealt{2014Sci...343.1330S}).

An estimate of the absolute accretion rate (i.e. in terms of mass accreted) can be obtained as $\dot{M}_\mathrm{BH} = k_\mathrm{bol} L_\mathrm{X}/(\eta c^{2})$ assuming a typical efficiency ($\eta$) of conversion of rest-mass into energy of 10\%. Taking $k_\mathrm{bol}=5$ yields $\dot{M}_\mathrm{BH} = 1.5 \times 10^{-5}$ M$\odot$ yr$^{-1}$, which is close to the Eddington accretion rate of a 1000 M$\odot$ BH ($2 \times 10^{-5}$ M$\odot$ yr$^{-1}$). The accretion rate is also one to two orders of magnitude higher than that of sub-Eddington to nearly Eddington accreting stellar-mass BHs with radio jet emission (e.g.  IC10 X-1, \citealt{2012ApJ...749...17C}; GRS1915+105, \citealt{2010A&A...524A..29R}), which favours an IMBH nature for NGC2276-3c. However, we note that it is also one order of magnitude smaller than that of the super-Eddington accreting galactic microquasar SS433 ($\dot{M}_\mathrm{BH} = 10^{-4}$ M$\odot$ yr$^{-1}$; \citealt{2004ASPRv..12....1F}).

\subsection{Jet kinematics}
The two VLA radio lobes have the same peak intensity (0.35 $\pm$ 0.06 mJy beam$^{-1}$ and 0.34 $\pm$ 0.06 mJy beam$^{-1}$; \citealt{2013MNRAS.436.3128M}) and distance to the compact source or X-ray position (arm-length ratio $R$=1.2). This argues against nebular radio emission and indicates that the VLA radio lobes most likely belong to a radio jet that is either not significantly affected by Doppler boosting effects or whose velocity vector of the emitting plasma is very close to the plane of the sky. From the arm-length ratio and the angle-to-line of sight ($\theta$), an estimate of the velocity in units of the speed of light ($\beta$) of a blob of plasma moving along the jet can be obtained from $\beta \mathrm{cos}\theta = (R-1)/(R+1)$ (\citealt{2014MNRAS.439L...1C}). Based on the lobe flux ratio and arm-length ratio, we assume $\theta$ = 80$^{\circ}$ and find $\beta$ = 0.4, while the constraint of cos $\theta \leq$ 1 yields $\beta \geq$ 0.1. Therefore the velocity of the radio jet is constrained to 2.3 $\times$ 10$^{4}$ -- 1.2 $\times$ 10$^{5}$ km s$^{-1}$. 

The localization of the IMBH in the spiral arm of a nearby galaxy suggests that it might have formed \textit{in situ}, as IMBHs of 10$^{2}$ -- 10$^{4}$ M$\odot$ are expected to form in dense young stellar clusters (\citealt{2009ApJ...694..302D}). However, the $M_\mathrm{BH}$ = 5 $\times$ 10$^{4}$ M$\odot$ of NGC2276-3c appears too large for that. The main alternative is that NGC2276-3c is the nucleus of an accreted stripped dwarf galaxy, as is likely the case for ESO 243-49 HLX-1 (\citealt{2012ApJ...747L..13F}; \citealt{2013ApJ...768L..22S}) and for the SMBH in the ultra-compact dwarf galaxy M60-UCD1 (\citealt{2014Natur.513..398S}). This is supported by the unusual morphology of the host spiral arm, its very high star formation rate (5--15 M$\odot$ yr$^{-1}$; \citealt{2015arXiv150101994W}) and the location of NGC 2276 in a galaxy group (\citealt{1997AJ....114..613D}). Interestingly, in this scenario the presence of the two nearby luminous ULXs is readily explained as a result of the star formation triggered by the passage of the stripped dwarf galaxy nucleus through the spiral arm. The symmetrical VLA radio jet lobe structure implies that NGC2276-3c must be travelling at a slower rate than it takes to inflate the lobes so that when passing through NGC 2276 the lobes retain their shape and are not distorted by the galaxy's interstellar medium (ISM). Typical time-scales for minor merger passages are $\sim$1 Gyr (e.g.  \citealt{2010MNRAS.404..575L}), with relative velocities between the two merging galaxies of 100--500 km s$^{-1}$ (e.g.  $\sim$200 km s$^{-1}$ for HLX-1; \citealt{2012MNRAS.423.1309M}). 
From the jet expansion velocity and considering a distance between the compact core and the outer VLA radio lobes of $\sim$322 pc, a range of dynamic jet ages of $\sim$2500 -- 14000 yr is obtained. This means that the time-scale for forming the lobes ($t_\mathrm{lobe}$) is much lower than the time-scale of a dwarf galaxy passing through NGC 2276 ($t_\mathrm{pass}$). The jet radio morphology is expected to be distorted or to present a non-linear shape during the course of a merger event (e.g.  as it is suggested in X-shaped radio galaxies; \citealt{2002Sci...297.1310M}; \citealt{2011A&A...527A..38M,2012A&A...544A..36M}) unless $t_\mathrm{lobe}<<t_\mathrm{pass}$. Hence, the finding that $t_\mathrm{lobe}<<t_\mathrm{pass}$ and that the jet velocity is much larger than the IMBH/dwarf galaxy's velocity favours the scenario in which the IMBH is the nucleus of a stripped dwarf galaxy, and might explain the unusual morphology and high star formation in the spiral arm of NGC 2276. 

\subsection{Jet power, efficiency and feedback}
An estimate of the total jet power, $Q_\mathrm{jet}$, can be derived from the radio luminosity of the compact jet. Using the correlation between core radio luminosity and kinetic power from \cite{2007MNRAS.381..589M}, which has the minimum scatter (0.4 dex) and is based on a sample of sub-Eddington accreting SMBHs with flat-spectrum compact jet cores, we obtain $Q_\mathrm{jet} = 4 \times 10^{40}$ erg s$^{-1}$, which is of the same order as the bolometric luminosity of NGC2276-3c ($\sim10^{40} -10^{41}$ erg s$^{-1}$). 

From the jet power and accretion rate, we can estimate the jet efficiency, or fraction of the accretion power that is used in the kinetic motion of the jet, $\eta_\mathrm{j}$ = $Q_\mathrm{jet}/(\eta\dot{M}_\mathrm{BH}c^{2}$), where $\eta$=10\%. Jet efficiencies of $\sim$10\% or higher (e.g.  \citealt{2014arXiv1406.7420N}) are estimated for massive radio galaxies, where the mechanical feedback from their jets produces cavities in their immediate inter-galactic environment (e.g.  \citealt{2007ARA&A..45..117M}). This mode of mechanical jet feedback (the so-called Ôradio modeÕ) in SMBHs is thought to be equivalent to the low/hard state of XRBs with low Eddington accretion ratios  ($<$ 2\%; e.g.  \citealt{2005MNRAS.363L..91C}). Using $Q_\mathrm{jet} = 4 \times 10^{40}$ erg s$^{-1}$ and $\dot{M}_\mathrm{BH} = 1.5 \times 10^{-5}$ M$\odot$ yr$^{-1}$, we obtain $\eta_\mathrm{j}$ = 46\%. The jet efficiency would be even higher if we e.g. take $k_\mathrm{bol}=1$ (this would yield $\dot{M}_\mathrm{BH} = 3 \times 10^{-6}$ M$\odot$ yr$^{-1}$ and $\eta_\mathrm{j}$ = 230\%) or use the relationship 
from \cite{2010ApJ...720.1066C}. 
In all cases, the estimated jet efficiency of NGC2276-3c is as large as that from massive radio galaxies with radio-mode mechanical feedback, demonstrating that the physical mechanisms of jet production and energy dissipation are similar not only for stellar-mass and SMBHs (\citealt{2012Sci...338.1445N}) but also for IMBHs. In SMBHs, such feedback is predicted to prevent star formation in the inner core of their host galaxy (radius $<$ 0.1 kpc; \citealt{2011ApJ...738...54K}) and to increase the proportion of stars deposited farther away from the centre (e.g.  \citealt{2013MNRAS.433.3297D}). This scenario is consistent with the optical image of NGC2276-3c (Fig.~\ref{fig1} b): a region of length $\sim$300 pc that is devoid of young stars is present in between the two peaks of the large-scale VLA radio lobes and oriented in the same direction as the parsec-scale EVN jet, while a larger population of stars is observed at the edges of the radio lobe emission. 
Randomly placing the VLA radio contours over other optical cavities observed in the \textit{Hubble Space Telescop} (\textit{HST}) image produces an obvious alignment in only 2 out of 19 trials, which makes a chance alignment between the optical cavity and the radio lobes unlikely (see also \citealt{2013MNRAS.436.3128M}).
Therefore, it is quite possible that the radio jet from the IMBH is clearing a cavity around it, in which new star formation is suppressed, while the star formation at the edges of this cavity could either take place when the material swept out by the jet collides with the ISM of NGC 2276 or have been triggered during the merger (e.g. \citealt{2012MNRAS.425L..46A}).

\section{Conclusions}
\label{conclusions}
Finding a large number of IMBHs is pivotal for understanding the role they play in SMBH/galaxy formation. The discovery of an IMBH in the spiral arm of NGC 2276 indicates that, in addition to the low-mass galaxies, the brightest ULXs in nearby spirals and the nuclei of minor mergers remain amongst the best IMBH candidates. The finding that the jet kinetic power of NGC2276-3c is comparable to its radiative output and that its jet efficiency is as large as those of radio galaxies indicates that, if seed BHs in the early Universe had powerful jets similar to that of NGC2276-3c, their jet mechanical feedback must be taken into account in cosmological simulations and studies of SMBH/galaxy growth. 

\section*{Acknowledgements}
MM acknowledges support from the Spanish Grant PNAYA2011-25527. TPR and ADS were funded as part of the STFC consolidated grants ST/K000861/1 and ST/L00075X/1.  The authors thank Eduardo Ros and Rodrigo Nemmen for insightful discussion. We thank the \textit{Chandra} X-ray Center for granting us a Director's Discretionary Time observation of NGC 2276. The EVN is a joint facility of European, Chinese, South African and other radio astronomy institutes funded by their national research councils.

\bibliographystyle{mn2e} 
\bibliography{referencesALL}

\label{lastpage}

\end{document}